\documentclass[a4paper,12pt]{article}
\usepackage{jheppub}
\usepackage[dvipsnames]{xcolor}
 \usepackage{hyperref}
\hypersetup{colorlinks=true,linkcolor=magenta,anchorcolor=green,citecolor=cyan,filecolor=black,menucolor=black,urlcolor=black}
\usepackage{stackrel}
\usepackage[compat=1.1.0]{tikz-feynman}
\usepackage{tikz,contour}
\usetikzlibrary{calc,decorations.markings}

\usepackage[force]{feynmp-auto}

\usepackage{bbm}
\usepackage{bbold}
\usepackage{float}

\definecolor{linkblue}{rgb}{0.1,0.3,.7}
\definecolor{forestgreen(web)}{rgb}{0.13, 0.55, 0.13}
\definecolor{lava}{rgb}{0.81, 0.06, 0.13}
\hypersetup{ 
breaklinks,
colorlinks,
citecolor=forestgreen(web),
filecolor=linkblue,
linkcolor=lava,
urlcolor=linkblue
}

\usepackage{caption}
\usepackage{subcaption}

\usepackage{pgfplots}
\usepackage[T1]{fontenc}
\usepackage[]{slashed}
\usepackage[]{bm}     
\usepackage{physics}
\usepackage{lipsum}
\usepackage{dsfont}
\usepackage{soul}

\usepackage{bbm,amsmath,graphicx,amssymb,amsfonts,amsthm}

\definecolor{bubbles}{rgb}{0.91, 1.0, 1.0}
\definecolor{aquamarine}{rgb}{0.5, 1.0, 0.83}
\definecolor{bubblegum}{rgb}{0.99, 0.76, 0.8}
\definecolor{blackbell}{rgb}{0.64, 0.64, 0.82}
\definecolor{dollarbill}{rgb}{0.72, 0.93, 0.6}

\pgfdeclarelayer{bg}    
\pgfsetlayers{bg,main}  

\definecolor{Mathematica}{HTML}{ed192d}

\usepackage{float}
\usepackage[T1]{fontenc} 

\begin{document}
\preprint{IPhT-t23/007}
\title{Bulk Landau Pole and Unitarity of Dual Conformal Field Theory}

\author{Ivo Sachs$^{a}$ and Pierre~Vanhove$^{b}$}
\affiliation{(a) Arnold-Sommerfeld-Center for Theoretical Physics, Ludwig-Maximilians-Universit\"at M\"unchen,
  Theresienstr. 37, D-80333 Munich, Germany}

\affiliation{(b) Institut de Physique Th\'eorique, Universit\'e  Paris-Saclay, CEA, CNRS, F-91191 Gif-sur-Yvette Cedex, France}

\date{\today}

\abstract{ The singlet sector of the $O(N)$ $\phi^4$-model in AdS$_4$ at  large-$N$, gives rise to a dual conformal field
  theory on the conformal boundary of AdS$_4$, which is a deformation
  of the generalized free field. We identify and compute an AdS$_4$ three-point one-loop  fish diagram that controls the exact large-$N$ dimensions and
  operator product coefficients (OPE) for all ``double trace''
  operators as a function of the renormalized $\phi^4$-couplings. We
  find that the space of $\phi^4$-coupling is compact with a boundary
  at the bulk Landau pole. The dual CFT is unitary only in an interval of {\it negative} couplings bounded by the Landau pole where the lowest OPE coefficient diverges. }

\maketitle

\section{Introduction}
To characterize an interacting quantum field theory in Minkowski
space-time we need to know the masses and spins of its asymptotic
states as well as the $S$-matrix elements between them. In curved
space-time there is no notion of $S$-matrix but in maximally symmetric
spaces such as de Sitter (dS$_4$) or anti-de Sitter (AdS$_4$)
space-time correlation functions evaluated on their conformal boundary
define a conformal field theory (CFT). Therefore, an interacting
quantum field theory in these spaces is characterized completely in
terms of the conformal dimensions of the primary fields of  that CFT
and their operator product expansion coefficients (OPE). This program
has been outlined in~\cite{Heemskerk:2009pn} and further explored in
many subsequent works,
including~\cite{Fitzpatrick:2011dm,Penedones:2010ue}.

In this note we consider a conformally coupled scalar field theory
with $\phi^4$ interaction in four-dimensional AdS. A free scalar in
AdS$_4$ with Dirichlet boundary conditions on its conformal boundary, is encoded in the CFT of a generalized free field~\cite{Heemskerk:2009pn} of conformal dimension $2$. The OPEs of the
latter have been determined in~\cite{Heemskerk:2009pn,Fitzpatrick:2011dm} by comparing the four-point
correlation function~\cite{Muck:1998rr} with the conformal block
expansion~\cite{Dolan:2000ut}. They give rise to double trace operators, in
  terminology analogous to  four-dimensional  Yang-Mills theory, which
  is holographically dual to string theory in AdS$_5\times S^5$. Bulk
  quantum field theories in AdS$_5$, being non-renormalizeable, are usually 
  defined as being ``the dual'' of a given boundary CFT. The present
  approach is the opposite: we construct a three-dimensional boundary CFT  for a  {\it given} renormalizeable bulk field theory in four-dimensional AdS.
The $\phi^4$ interaction does not affect the
spectrum of the CFT in perturbation theory, but the dimensions and OPE
coefficients of  double trace operators are
corrected~\cite{Heemskerk:2009pn,Fitzpatrick:2011dm}.
The one-loop
correction to the bulk correlations function and thereby the CFT data
was then found in~\cite{Bertan:2018afl,Bertan:2018khc,Heckelbacher:2022fbx}. The
calculation of the loop integrals as well as the conformal block
expansion at this order is rather exhaustive,  but an extension to
higher loops is not an easy task (see~\cite{Heckelbacher:2022fbx} for
a discussion). On the other hand it is well-known that in Minkowski
space an all loop extension is available in the large-$N$ limit of the
${\lambda\over N}(\phi_i\phi^i)^2$ theory or $O(N)$  model (e.g.~\cite{Moshe:2003xn} for a review). The leading
large-$N$ contribution of the four-point function is given by the sum
of a necklace of  multi-bubble
diagrams which, thanks to momentum conservation, are just the power of
the one-loop bubble. In the large-$N$ limit the perturbative
corrections can be summed into
\begin{align}\label{eq:B_f}
    W(p)\sim \lambda\frac{\delta^{(4)}(p_1+\dots+p_4)}{1- B(|p_1+p_2|)},
\end{align}
where $B(p)$ is the one-loop four-point bubble
contribution. However, it was shown in~\cite{Coleman:1974jh} that this  features a tachyonic mode in large-$N$ limit. We identify a manifestation of that pathology in
AdS$_4$. This requires a resummation of the multi-loop bubble diagrams in
AdS$_4$ which has so far been elusive. Here we solve this problem and  derive the associated renormalized spectral function $\mathcal B_{\rm ren}(\nu)$ in eq.~\eqref{e:Breno1loop}.

\section{Tree-Level Diagrams}
Quite generally, the relation of AdS$_4$ boundary four-point functions
$W(\vec{x}_i)$, to correlators of a three-dimensional CFT primary field
$\mathcal{O}_2(\vec{x})$ of dimension 2 is given by   
\begin{align}	\label{eq:generalized_free_field}
\langle \mathcal{O}_2(\vec{x}_1) \cdots \mathcal{ O}_2(\vec{x}_4)\rangle&=W(\vec{x}_1,\vec{x}_2,\vec{x}_3,\vec{x}_4 )
\end{align}
We consider the Poincar\'e patch with coordinates $X=\{\vec{x},z\}\in \mathbb R^3\times \mathbb R_+$. The CFT four-point function has an expansion in terms of conformal blocks as

\begin{equation}\label{eq:N_c_23_0_p}
    W(\vec{x}_1, \vec{x}_2, \vec{x}_3,
  \vec{x}_4)=\int_{\mathbb{R}} \frac{\dd\nu}{2\pi} D(\nu)
              g(\vec{x}_1,\vec{x}_2,\vec{x}_3,\vec{x}_4; \nu )
\end{equation}
where 
 \begin{equation}\label{eq:g}
 g(\vec{x}_i;\nu )\propto \int\limits_{\partial AdS_4} \dd^3\vec
    x  \langle \mathcal{O}_2(\vec{x}_1) \mathcal{O}_2(\vec{x}_2)\mathcal{O}_{\nu}(\vec{x})\rangle\langle \tilde{\mathcal{O}}_{-\nu}(\vec{x}){\mathcal{O}}_2(\vec{x}_3) {\mathcal{O}}_2(\vec{x}_4)\rangle
\end{equation}
encodes the contribution of an internal double trace 
operator $\mathcal{O}_{\nu}(\vec{x})$  (and its descendants) of conformal dimension
$\Delta(\nu)=\frac{3}{2}+i\nu$, and $\tilde{\mathcal{O}}_{-\nu}$ is the ``shadow'' primary field of
dimension $3-$dim$[\mathcal{O}_\nu]=3/2-i\nu$ (see for instance~\cite[\S2]{Meltzer:2019nbs}).
The dimension of the internal
primaries are then given by the poles of the spectral function
$D(\nu)$ while the OPE coefficients into the double trace
operators are encoded in the residues of the latter. Conversely,  
$D(\nu)$ is obtained by integrating $W(\vec{x}_i)$
against a CFT three-point function
\begin{equation}\label{eq:3ptcft}
   \langle \mathcal{O}_{\Delta}(\vec{x}_1) \mathcal{ O}_{\Delta}(\vec{x}_2)\mathcal{O}_{\nu}(\vec{x}_0)\rangle=    \frac{1}{x_{12}^{2\Delta-\Delta(\nu)} x_{10}^{\Delta(\nu)} x_{20}^{\Delta(\nu)}}
\end{equation}
which satisfies an orthogonality relation
\begin{multline}\label{e:ortho_1_p}
    \int\limits_{\partial \text{AdS}_4} \dd^3 x_1  \dd^3 x_2  \langle \mathcal{O}_{\nu}(\vec{x})\mathcal{O}_2(\vec{x}_1) \mathcal{O}_2(\vec{x}_2)\rangle\langle \tilde{\mathcal{O}}_1(\vec{x}_1) \tilde{\mathcal{O}}_1(\vec{x}_2)\mathcal{O}_{-\nu'}(\vec{x}')\rangle \cr
= \frac{4\pi^4|\Gamma(i\nu)|^2}{|\Gamma(\frac{3}{2}+i\nu)|^2}\delta(\nu-\nu')\delta(\vec{x}-\vec{x}')\,.
\end{multline}

\noindent\textbf{Disconnected four-point function:} Dismissing the identity conformal block we have 
\begin{equation}	\label{eq:generalized_free_field_2}
W^{(\textrm{disc})}(\vec{x}_1,\vec{x}_2,\vec{x}_3,\vec{x}_4 )=
		\frac{1}{(x^2_{13}x^2_{24})^{2}}+ \frac{1}{(x^2_{14}x^2_{23})^{2}}\,.
\end{equation}
The result for the spectral function was given in~\cite{Heemskerk:2009pn,Fitzpatrick:2011dm}
\begin{equation}\label{eq:mean_p}
    D^{(\textrm{disc})}(\nu)=\frac{2\pi^{\frac{3}{2}}|\Gamma(\frac{5}{4}+i\frac{\nu}{2})|^2\Gamma(\frac{3}{2}-i\nu)\Gamma(\frac{3}{4}+i\frac{\nu}{2})^2 }{|\Gamma(\frac{1}{4}+i\frac{\nu}{2})|^2\Gamma(i\nu)\Gamma(\frac{3}{4}-i\frac{\nu}{2})^2}\,,
\end{equation}
whose simple poles and their residues imply the (mean field) double
trace dimensions and OPE's as in~\cite{Heemskerk:2009pn}
\begin{equation}\label{eq:mean_c}
    \nu_n=-i\left(\frac{5}{2}+2n\right)\;;  \qquad  |\bar c(\nu_n)|^2=i\text{Res}D^{(\textrm{disc})}(\nu_n)\,.
\end{equation}

\begin{figure}[t]
                \begin{center}
\includegraphics[scale=1.5]{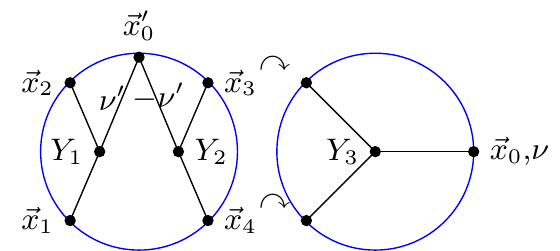}
                          \caption{Sketch of the integration of the bulk four-point function left  (with resolved $\delta$-function vertex as in~\eqref{eq:complete_p}) against the three-point function with two shadow operators $\langle \tilde{\mathcal{O}}_1(\vec{x}_3) \tilde{\mathcal{O}}_1(\vec{x}_4)\mathcal{O}_{-\nu}(\vec{x})\rangle $ on the right  to extract  $D(\nu)$.}
                        \label{fig:Dnu}
                \end{center}
        \end{figure}

\noindent\textbf{Cross diagram:}  Here we evaluate the spectral function using a different  method that will be useful at loop level. As represented in the diagram on the left of fig.~\ref{fig:Dnu}, we resolve the bulk four-point vertex  using the representation of the delta-function~\cite{Meltzer:2019nbs}
\begin{equation}\label{eq:complete_p}
    \delta(Y_1-Y_2)=\int\limits_{-\infty}^\infty
  \frac{\dd\nu \nu^2}{\pi} \int\limits_{\partial\text{AdS}_4}
  \dd^3 x \bar\Lambda_{\Delta(\nu)}(\vec{x},Y_1)\bar\Lambda_{\Delta(-\nu)}(\vec{x},Y_2)
\end{equation}
in terms of  the bulk-to-boundary propagator 
\begin{equation}\label{e:LambdaBb}
    \bar\Lambda_\Delta(\vec{x},X')=\frac{n_\Delta z'^{\Delta}}{\left((\vec{x}'-\vec{x})^2+z'^2\right)^{\Delta}}\,,\qquad {n_\Delta}= \frac{\Gamma\left(\Delta\right)}{2\pi^{\frac{3}{2}}\Gamma(\Delta-\frac12)}.
\end{equation}
 for a scalar of mass
$m^2=-\frac{9}{4}-\nu^2$ and dimension $\Delta(\nu)=\frac32+i\nu$. 
In conjunction with~\eqref{eq:N_c_23_0_p}-\eqref{eq:g} this shows that the tree-level spectral function for the cross diagram
$  D^{(\times)}(\nu) $ is given by (the coupling constant $\lambda$ times) the relative normalizations of the CFT three-point correlator entering in~\eqref{e:ortho_1_p} and the bulk three-point function 
 \begin{align}
  \label{e:W0}
  W_\nu^{(0)}(\vec x_1,\vec x_2,\vec x_0)&= \int\limits_{\textrm{AdS}_4}
  \dd Y \bar\Lambda_{2}(\vec{x}_1,Y)\bar\Lambda_{2}(\vec{x}_2,Y)\bar\Lambda_{\Delta(\nu)}(\vec{x}_0,Y)\cr
  &={|\Gamma(\frac{5}{4}+i\frac{\nu}{2})|^2 \Gamma(\frac{3}{4}+i\frac{\nu}{2})^2\over 4\pi^4\Gamma(1+i\nu)}{1\over x_{12}^{4-\Delta(\nu)} x_{10}^{\Delta(\nu)} x_{20}^{\Delta(\nu)}}\,.
\end{align}
Therefore
\begin{equation}\label{eq:dx_p}
    D^{(\times)}(\nu)=-\frac{\lambda}{2\pi^{5/2}}  \frac{|\Gamma(\frac{5}{4}+i\frac{\nu}{2})|^4 \Gamma(\frac{3}{4}+i\frac{\nu}{2})^4}{\Gamma(i\nu)\Gamma(\frac{3}{2}+i\nu)}\,,
\end{equation}
which now has double poles. Comparing this with the perturbative series of the conformal block expansion~\eqref{eq:N_c_23_0_p}
\begin{multline}
   W(\vec{x}_1,\dots,\vec{x_4}) =\sum\limits_{n}|\bar c_{\nu(n)}|^2
    g(\vec{x}_i; \nu_n)
  -i\sum\limits_{n}|\bar c_{\nu(n)}|^2\gamma^{(1)}_n \partial_\nu g(\vec{x}_i;\nu)|_{\nu_n}\cr
  +\sum\limits_{n} (|\bar c|^2_{\nu(n)})^{(1)}\gamma^{(1)}_ng(\vec{x}_i;\nu_n)\,,
\end{multline}
one then identifies the first order double trace anomalous dimensions $\gamma^{(1)}_n$ and squared OPE's $(|c(\nu_n)|^2)^{(1)}$ as 
\begin{align}\label{eq:gamma_1_p}
 \gamma^{(1)}_n=i\frac{\text{Res}D^{(\times)}(\nu_n)}{\text{Res}D^{(\textrm{disc})}(\nu_n)}=\frac{\lambda}{16\pi^2}
\end{align}
and
\begin{align}\label{eq:coeff_1_p}
  (|c(\nu_n)|^2)^{(1)}=\frac{\partial\text{Res}D^{(\textrm{disc})}(\nu_n)}{\partial{\nu_n}}=\frac{1}{2}\frac{\partial|\bar c(\nu_n)|^2}{\partial{n}}\,,
\end{align}
thus reproducing the expression given in~\cite{Fitzpatrick:2011dm}.

\begin{figure}[t]
  \begin{center}
    \includegraphics[scale=1.5]{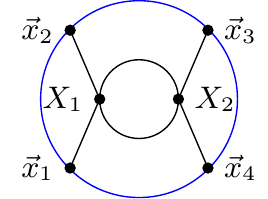}
                          \caption{$s$-channel one-loop bubble }
                        \label{fig:1_4pt_diagram}
                \end{center}
        \end{figure}

        \section{One-loop Diagram}

\begin{figure}[t]
                \begin{center}
 \includegraphics[scale=1.5]{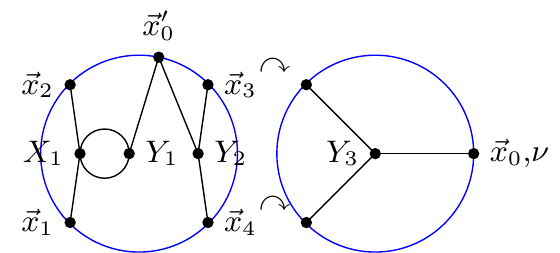}
                        \caption{Sketch of the reduction of the
                          one-loop bubble in fig.~\ref{fig:1_4pt_diagram} to the fish diagram of fig.~\ref{fig:fish}.}
                        \label{fig:BubbleToFish}
                \end{center}
        \end{figure}

\begin{figure}[b]
                \begin{center}
\includegraphics[scale=1.5]{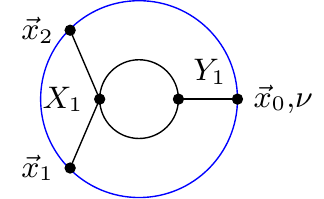}
                        \caption{Three point fish diagram}
                        \label{fig:fish}
                \end{center}
        \end{figure}

A key observation for determining  the spectral
function directly for the one-loop amplitude in fig.~\ref{fig:1_4pt_diagram} is to use the delta-function~\eqref{eq:complete_p} to factorize the correlator into three-point functions as in fig.~\ref{fig:BubbleToFish}
and use the orthogonality 
relation~\eqref{e:ortho_1_p} to express the one-loop graph spectral
function in term of the fish three-point function $W^{(\rangle\!\bigcirc\!-)}_\nu(\vec{x}_1,\vec{x}_2,\vec{x}_0)$  of fig.~\ref{fig:fish}
\begin{equation}\label{eq:B-c}
  D^{(\rangle\!\bigcirc\!\langle)}(\nu)= D^{(\times)}(\nu) \times\frac{W^{(\rangle\!\bigcirc\!-)}_\nu(\vec{x}_1,\vec{x}_2,\vec{x}_0) }{ W^{(0)}_\nu(\vec{x}_1,\vec{x}_2,\vec{x}_0)}.
\end{equation}

The fish diagram has the  ultraviolet divergence for colliding bulk
points familiar from flat space which we regulate using the AdS-invariant cut-off introduced
in~\cite{Bertan:2018afl}. This amounts to modify the bulk-to-bulk propagator as
\begin{equation}\label{e:LambdaDelta}
    \Lambda^\delta(X,X')= {4z^2z'^2\over (1+\delta)^2((\vec x-\vec{x}')^2+z^2+z'^2)^2-4z^2z'^2}
\end{equation}
so that it is finite for coincident bulk points,  $\Lambda^\delta(X,X)={1\over \delta (\delta+2)}$,  $\delta>0$. 
The $X_1$ integral then proceeds as in~\cite[\S4.2.2]{Heckelbacher:2022fbx}  to give
\begin{align}
 \label{eq:fish_nu_p}
		W_{
                  \nu}^{(\rangle\!\bigcirc\!-)}(\vec{x}_1,\vec{x}_2,\vec{x}_0)=&\frac{-\lambda^2}{32\pi^2}\frac{n_\Delta^2 n_{\Delta(\nu)}}{ x_{10}^{\Delta(\nu)}x_{20}^{\Delta(\nu)}
                  x_{12}^{4-\Delta(\nu)}}\\
  &            \times    \int\limits_{\mathbb{R}_+^4}{\dd^4
                Xz^{\Delta(\nu)}\over |X|^4 |X-\vec w_2|^4 }  \left(\log2\delta+2+\log\left(\frac{z^2  |\vec{w}_2|^2}{|X|^2|\vec w_2-X|^2}
		\right)\right).\nonumber
	\end{align}
where $\vec w_2=\vec x_2/|\vec x_2|$. It is then clear from rotation invariance that the integral can only depend on $\nu$. Furthermore, the ultraviolet divergence from the collapsing loop in fig.~\ref{fig:fish} is  proportional the three point function~\eqref{e:W0}.   The renomalization scheme of~\cite{Bertan:2018afl,Heckelbacher:2022fbx} amounts to subtracting from $W_{\nu}^{(\rangle\!\bigcirc\!-)}(\vec{x}_i)$
the counter-term
$\frac{\lambda^2}{32\pi^2}(\log(\delta/2)+\frac{11}{3})\,W_{\nu}^{(0)}(\vec{x}_i)$.  The logarithmic term in the integrand is conveniently written as a $\nu$-derivative. Then, making use of~\eqref{e:W0} we can read off the renormalised one-loop spectral function 
\begin{equation}\label{e:Dreno1loop}
 D_{\rm ren}^{(\rangle\!\bigcirc\!\langle)}(\nu)= \frac{\lambda_R}{32\pi^2}\Big(\log(4)-\frac{5}{3}+\psi\left(\frac{5}{4}+i\frac{\nu}{2}\right)+\psi\left(\frac{5}{4}-i\frac{\nu}{2}\right)-2\psi(2)\Big)D^{(\times)}(\nu)\,,
\end{equation}
where $\psi(x)=\Gamma'(x)/\Gamma(x)$ is the digamma function
and the renomalized coupling constant satisfies
\begin{equation}\label{e:LambdaR}
    \frac{16\pi^2}{\lambda}=\frac{16\pi^2}{\lambda_R}+\frac{1}{2}\left(\log(
\frac{\delta}{2})+\frac{11}{3}\right)\,.
\end{equation}
The function~\eqref{e:Dreno1loop}, together with $D^{(\times)}(\nu)$ provides closed
expressions for all one-loop $s$-channel anomalous dimensions and
OPE's, which can be checked to agree numerically with the ones
previously obtained
in~\cite{Bertan:2018khc,Heckelbacher:2022fbx}. Note
that~\eqref{e:Dreno1loop} is structurally similar to the spectral
function obtained previously in~\cite{Carmi:2019ocp} for AdS$_2$ and
AdS$_3$, ingeniously using the bootstrap approach. However, we will
see that the physics derived form~\eqref{e:Dreno1loop} is rather
different.

\section{Large-$N$ $O(N)$ Model}
We now consider the curved space version of the vector  model where the
scalar field $\phi$ transforms in the fundamental representation of
$O(N)$, together with the ${\lambda\over 4N} (\phi_i\phi^i )^2$ interaction. We will focus on the CFT data encoding the
singlet sector. Thus, we  consider the large-$N$
limit of the conformal block expansion of the singlet  four-point function 
\begin{equation}\label{eq:sing-4}
 {1\over N^2}  \langle \mathcal{O}_i(\vec{x}_1)
   \mathcal{O}^i(\vec{x}_2)\mathcal{O}_p(\vec{x}_3) \mathcal{
     O}^p(\vec{x}_4)\rangle =\int\limits_{\mathbb R} {\dd \nu\over2\pi} D(\nu) c_{ii}
  c_{pp} g(\vec x_1,\dots,\vec x_4,\nu)+O(1/N^2).
 \end{equation}
 The identity OPE is $O(1)$ in the large-$N$ limit, while
the disconnected contribution to the double trace dimension $\Delta$
and (squared) OPE's $|\bar c|^2$ is again given by~\eqref{eq:mean_p} after re-scaling by $1/N$.
For the cross diagram with interaction vertex 
\begin{equation}
    \frac{\lambda}{N}\left(\delta_{ij}\delta_{pq}+\delta_{ip}\delta_{jq}+\delta_{iq}\delta_{jp}\right)\,,
\end{equation}
 the first term in the bracket dominates, at large $N$, for the singlet correlator~\eqref{eq:sing-4}.  This then results in~\eqref{eq:gamma_1_p} and~\eqref{eq:coeff_1_p} for the anomalous dimension and OPE, after re-scaling $\text{Res}D^{(\textrm{disc})}$ and $\text{Res}D^{(\times)}$ both with $1/N$. 
At one-loop and large-$N$ the $s$-channel in
fig.~\ref{fig:1_4pt_diagram}  dominates and by consequence the
$D^{(\rangle\!\bigcirc\!\langle)}_{\rm ren}(\nu)$ is again given by~\eqref{e:Dreno1loop} rescaled by $1/N$. 


\medskip

\begin{figure}[t]
  \begin{center}
    \includegraphics[scale=1.5]{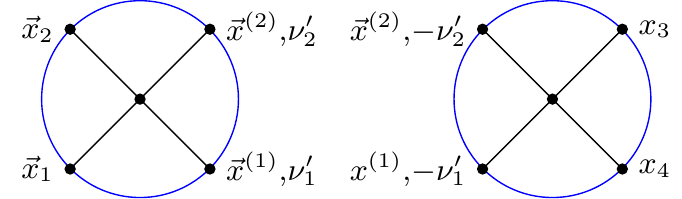}
                             \caption{The split representation of  $W^{(\rangle\!\bigcirc\!\langle)}(\vec x_1,\dots,\vec
                          x_4)$ in fig.~\ref{fig:1_4pt_diagram}
                          as a product of two cross diagrams.}
                        \label{fig:W1split}
                \end{center}
        \end{figure}

 To see the resummation of the $s$-channel bubble diagrams we first give an
  alternative representation  of the bubble diagram  in fig.~\ref{fig:1_4pt_diagram}  as a product of two cross diagrams, using the split representation ~\cite{Costa:2014kfa} 
depicted in fig.~\ref{fig:W1split}. However, we need to modify this representation to treat the aforementioned ultraviolet divergence for coincident bulk points. The regularized bulk-to-bulk propagator has the 
split representation
\begin{equation}
   \Lambda^\delta(X_1,X_2) =\int\limits_{-\infty}^{+\infty} {\nu^2d\nu\over\pi}
   \int\limits_{\partial AdS_4} {\dd^3 x \bar\Lambda_{\nu}(\vec x,X_1)\bar\Lambda_{-\nu}(\vec x,X_2)\over
     \nu^2+\frac{1}{4}+f(\nu,\delta)
     },\end{equation}
     where $f(\nu,\delta)$ can be chosen in such a way as to reproduce the $\delta$-regularization introduced in~\eqref{e:LambdaDelta}.
The ultraviolet divergences arise from the large-$\nu$ behaviour are then regulated by the $\delta$ regularisation. The precise
  choice of regulation is not important here as long as it preserves
  AdS-invariance. For instance, a cut-off regulator on the integral or
 zeta-function regularisation as in~\cite{Giombi:2013yva} could also
 be used.

After adding the counter term 
 restoring the factor $\frac{\lambda_R^2}{2}$, the one-loop diagram in
 fig.~\ref{fig:1_4pt_diagram} then takes the form 
\begin{equation}\label{eq:4ptr}
   W_{\rm ren}^{(\rangle\!\bigcirc\!\langle)}(\vec{x}_i
   )=   \frac{\lambda_R}{2} \int_{\mathbb R} \dd\nu
   D^{(\times)}(\nu) \mathcal B_{\rm ren}(\nu)
   g(\vec{x}_i;\nu) .
\end{equation}
For convenience we expressed the one-loop spectral function~\eqref{e:Dreno1loop} as the  product of $D^{(\times)}(\nu)$ and the one-loop bubble function
\begin{equation}\label{e:Breno1loop}
 \mathcal{B}_{\rm ren}(\nu)= \frac{\lambda_R}{32\pi^2}\Big(\log(4)-\frac{5}{3}+\psi\left(\frac{5}{4}+i\frac{\nu}{2}\right)+\psi\left(\frac{5}{4}-i\frac{\nu}{2}\right)-2\psi(2)\Big)\,.
\end{equation}
The advantage of the split representation is that it straight forwardly extends to the two-loop diagram. It is not hard to see that the two-loop necklace
 contribution $W^{(2)}(\vec x_1,\dots,\vec x_4)$ from~\eqref{eq:4ptr} by replacing 
$\mathcal
B_{\rm ren}(\nu)$ by $(\mathcal
B_{\rm ren}(\nu))^2$. In this way, the multi-bubble diagrams can be summed up resulting in 

\begin{align}\label{eq:resum4pt}
     W_{\rm ren}(\vec{x}_1,\dots,\vec{x}_4
  )&:=\sum\limits_{k=1}^\infty W^{(k)}_{\rm ren}(\vec{x}_1,\dots,\vec{x}_4 )\\
\nonumber    &=\int\limits_{-\infty}^{+\infty} \frac{\dd
               \nu}{2\pi}\frac{D^{(\times)}(\nu)}{1-{\mathcal{B}}_{\rm
               ren}(\nu)}g(\vec{x}_1,\dots,\vec{x}_4;\nu).
\end{align}
 In the  flat space limit $\nu\propto |p|\to
\infty$ where $p$ is the four dimensional momentum,  $\mathcal
B_{\rm ren}(\nu)$ in the integrand of~\eqref{eq:resum4pt} gives the generalization of flat space
expression in~\eqref{eq:B_f}. This will be discussed further below when
comparing with the flat space results of~\cite{Coleman:1974jh}.
The disadvantage of the split representation is however that a closed expression form of $\mathcal
B_{\rm ren}(\nu)$ in the Mellin representation is not available. Luckily we don't need it since we already evaluated $\mathcal
B_{\rm ren}(\nu)$ in a different way in~\eqref{e:Dreno1loop}.

\begin{figure}[t]
  \centering
  \subcaptionbox{}{
    \includegraphics[scale=1.5]{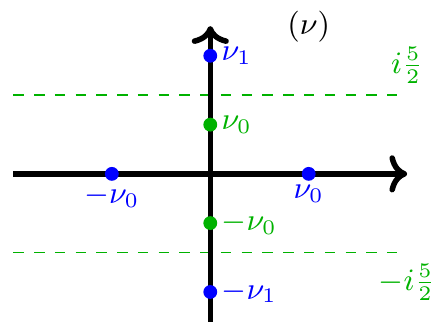}
  }
  \subcaptionbox{  }{
    \includegraphics[scale=1.5]{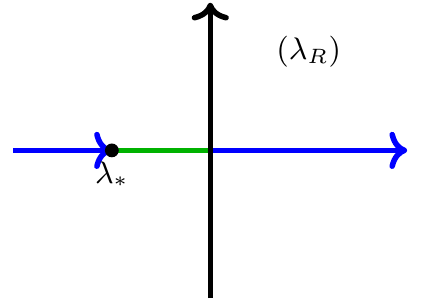} }
  \caption{Fig.~(a) gives the solutions for $\nu$ of~eq.~\eqref{eq:B=L} 
    according the  range of $\lambda_R$ in fig.~(b). $\lambda_*$ is the Landau pole. }                      \label{fig:landauPole}
        \end{figure}

Thanks to the fall-off of the conformal block~\cite{Ponomarev:2019ofr}, the contour can be closed in the lower half $\nu$-plane.
There are two types of poles that contribute to the
integral~\eqref{eq:resum4pt}: i) The double poles of
$D^{(\times)}(\nu)$ in eq.~\eqref{eq:dx_p} and ii) The
zeroes of $1-\mathcal{ B}_{\rm ren}(\nu)$. The set i)
coincides with the poles $\nu_n$ of $\mathcal{ B}_{\rm ren}(\nu)$, so that, using~\eqref{eq:gamma_1_p}, the integral has just simple poles  of $O(\lambda_R^0)$. Furthermore,  
\begin{equation}
\stackrel[\nu=\nu_n]{}{\textrm{Res}}\left( \frac{iD^{(\times)}(\nu)}{1-{\mathcal{B}}_{\rm
               ren}(\nu)}\right)={1\over \gamma^{(1)}} \stackrel[\nu=\nu_n]{}{\textrm{Res}}((\nu-\nu_n) D^{(\times)}(\nu))=-|\bar c(\nu_n)|^2 
\end{equation}
and thus does cancel the disconnected (mean field)
contribution. This is required by consistency since otherwise the mean field double trace operators would continue to contribute at finite coupling $\lambda$. Concerning ii) we need to
solve the equation $\mathcal{ B}_{\rm ren}(\nu)=1$ or, equivalently (see also~\cite{Carmi:2018qzm,Carmi:2019ocp})
\begin{equation}\label{eq:B=L}
    \log(4)-\frac{5}{3}+\psi\left(\frac{5}{4}+i\frac{\nu}{2}\right)+ \psi\left(\frac{5}{4}-i\frac{\nu}{2}\right)-2\psi(2)=\frac{32\pi^2}{\lambda_R}\,.
  \end{equation}
For $\lambda_R=0$ the solutions of~\eqref{eq:B=L} are given by the poles of $\psi\left(\frac{5}{4}+i\frac{\nu}{2}\right)$ which correspond the double trace dimensions in the mean field theory. Then, increasing $\lambda_R$, the double trace dimensions grow in accordance with perturbation theory  approaching a finite value at 
for $\frac{1}{\lambda_R}\to 0_+$. This function can then be continuously be extended to negative
values of $\frac{1}{\lambda_R}$ down to the critical coupling
  \begin{equation}\label{e:lambdaP}
    {32\pi^2\over\lambda_*}=\frac{13}{3}-4 \ln \! \left(2\right)-\pi\simeq-1.58
  \end{equation}
where eq.~\eqref{eq:B=L} is solved for 
$\nu=0$. At this point a new double trace operator of dimensions $\Delta(0)={3\over2}$ 
appears.
All this is 
represented in
fig.~\ref{fig:landauPole}. For
 $\lambda_*\leq \lambda_R\leq 0$ this solution $\nu_0$ (in green) moves down the negative imaginary axis with the dimensions of the first double trace operator covering the interval ${3\over2}\leq\Delta(\nu)\leq 4$ approaching again the mean field value $\Delta=4$. The space of couplings is therefore compact. 

However, analysing the solutions near $\lambda_*$ reveals an additional set of poles.\footnote{We would like to thank Shota
  Komatsu for pointing out the existence of these additional poles and their
  possible interpretation.} Indeed, 
when $\lambda_R\leq \lambda_*$ or $\lambda_R\geq 0$  there are  additional solutions
with 
$\nu$ real (in blue), associated with non-unitary  operators of complex dimensions $\Delta(\nu)$. We then claim this to be the manifestation in AdS of the tachyonic pathology in the flat space analysis
of~\cite{Coleman:1974jh} at $\lambda_R>0$.
To support of this interpretation we note that in the flat space limit $\nu\to\infty$, the
spectral function behaves as $\mathcal B_{\rm ren}(\nu)\sim
{\lambda_R\over 32\pi^2} \log(\nu)$ in accordance with~\cite[\S~III.C]{Coleman:1974jh}. On the other hand, for negative coupling,  $\lambda_*\leq \lambda_R\leq 0$ the dual CFT is unitary, which comes as a surprise given the instability of the potential in flat space for negative coupling.

It is also interesting to compare with  the spectral functions for AdS$_2$
and AdS$_3$  determined in~\cite{Carmi:2018qzm} which have only purely
imaginary roots associated to operators of real dimensions, in agreement
with the flat space limit where no 
tachyons are expected in two and three dimensions~\cite{Coleman:1974jh}.  

The non-perturbative OPE coefficient for the $n$-th double trace operator is in turn given by 
\begin{equation}
  |c(\nu_n)|^2=  \stackrel[\nu=\nu_n]{}{\text{Res}}\left(\frac{i D^{(\times)}(\nu)}{1-{\mathcal{B}_{\rm
      ren}}(\nu)}\right).
\end{equation}
For $\nu_0\to 0_+$,  $|c(\nu_0)|^2$ grows without bound. We interpret this feature as a manifestation in CFT of the Landau pole of the $\phi^4$ theory at negative coupling: The generalized free field represented by a free scalar in AdS$_4$ is one point in a continuous family of CFT's which can be parametrized by the bulk coupling $\lambda_R$ or, equivalently by the anomalous dimension $\gamma^{(1)}$ in eq.~\eqref{eq:gamma_1_p}. However, when $\gamma^{(1)}=\frac{\lambda_*}{32\pi^2}$, the spectral representation~\eqref{eq:N_c_23_0_p} develops a singularity since the lowest pole crosses the integral  contour. This is reminiscent of the Landau pole in $\phi^4$ theory which leads to a divergence of loop integrals~\cite{Landau:1954cxv}. Note that while the $\lambda_R$-dependence of the double trace dimensions and OPE's is renormalization scheme dependent the relation between OPE and double trace dimension is not. In CFT one can resolve this singular point by moving the $\nu$ integration contour to the left of this pole and adding the conformal block of this operator by hand, similar to the identity conformal block which was similarly not included in~\eqref{eq:generalized_free_field} and~\eqref{eq:N_c_23_0_p}. For the same reason this block has to be removed for the integral in fig.~\ref{fig:Dnu} to converge.  See~\cite{Caron-Huot:2017vep} for a more detailed discussion. 

\section{Conclusion}
A key result in this letter is the renomalized spectral function in eq.~\eqref{e:Breno1loop} which contains all relevant information of the CFT representation of the large-$N$, $O(N)$-model in AdS$_4$. 
 Other methods have been used for determining the spectral function
 $\mathcal B_{\rm ren}(\nu)$. For instance, in~\cite{Carmi:2018qzm} the spectral
function has been obtained for AdS$_{d+1}$ with $d<3$, but the result is not valid for  $d\geq3$ due to the one-loop ultraviolet divergences. Thanks to the AdS invariant regulator~\cite{Bertan:2018afl,Heckelbacher:2020nue} the renormalized spectral function is fully determined by the regulated one-loop fish diagram in fig.~\ref{fig:fish} for which we give a closed form expression in eq.~\eqref{eq:fish_nu_p}.

After extracting the dimensions of the double trace operators and the
OPE's for this spectral function we then found that 
$\lambda_R=\infty$ is not a singular point. However, approaching the
Landau pole of the $O(N)$ model in the negative coupling
regime a double trace operator  of dimension $3/2$ develops a complex dimension which
results in a singularity in the spectral representation. The
present analysis then shows the appearance of complex dimension operators that  translate into a non-unitary CFT everywhere outside the green interval in fig.~\ref{fig:landauPole}(b), 
which covers all positive $\lambda_R$, 
in accordance with the tachyonic mode found in~\cite{Coleman:1974jh}.  
For a negative coupling $\lambda_R\in [\lambda_*,0]$ the theory is unitary, and the space of couplings is compact. In closing, let us mention that for $\lambda_R=
\lambda_*$ there is a marginal ``quadruple trace'' operator in the OPE of two double trace operators. It would be interesting to investigate its effect on the CFT and on the bulk theory upon giving  to this operator a vacuum expectation value.

\acknowledgments
  We thank Till Heckelbacher, Igor Klebanov, Shota Komatsu, Juan
  Maldacena, Zhenya Skvortsov,   Pedro Vieira,  for discussions.  I.S. is supported by the Excellence Cluster Origins of the DFG under Germany’s Excellence Strategy EXC-2094 390783311.
 The research of P.V. has received funding from the ANR grant ``SMAGP''
ANR-20-CE40-0026-01.


\begin{thebibliography}{20}%
\makeatletter
\providecommand \@ifxundefined [1]{%
 \@ifx{#1\undefined}
}%
\providecommand \@ifnum [1]{%
 \ifnum #1\expandafter \@firstoftwo
 \else \expandafter \@secondoftwo
 \fi
}%
\providecommand \@ifx [1]{%
 \ifx #1\expandafter \@firstoftwo
 \else \expandafter \@secondoftwo
 \fi
}%
\providecommand \natexlab [1]{#1}%
\providecommand \enquote  [1]{``#1''}%
\providecommand \bibnamefont  [1]{#1}%
\providecommand \bibfnamefont [1]{#1}%
\providecommand \citenamefont [1]{#1}%
\providecommand \href@noop [0]{\@secondoftwo}%
\providecommand \href [0]{\begingroup \@sanitize@url \@href}%
\providecommand \@href[1]{\@@startlink{#1}\@@href}%
\providecommand \@@href[1]{\endgroup#1\@@endlink}%
\providecommand \@sanitize@url [0]{\catcode `\\12\catcode `\$12\catcode
  `\&12\catcode `\#12\catcode `\^12\catcode `\_12\catcode `\%12\relax}%
\providecommand \@@startlink[1]{}%
\providecommand \@@endlink[0]{}%
\providecommand \url  [0]{\begingroup\@sanitize@url \@url }%
\providecommand \@url [1]{\endgroup\@href {#1}{\urlprefix }}%
\providecommand \urlprefix  [0]{URL }%
\providecommand \Eprint [0]{\href }%
\providecommand \doibase [0]{http://dx.doi.org/}%
\providecommand \selectlanguage [0]{\@gobble}%
\providecommand \bibinfo  [0]{\@secondoftwo}%
\providecommand \bibfield  [0]{\@secondoftwo}%
\providecommand \translation [1]{[#1]}%
\providecommand \BibitemOpen [0]{}%
\providecommand \bibitemStop [0]{}%
\providecommand \bibitemNoStop [0]{.\EOS\space}%
\providecommand \EOS [0]{\spacefactor3000\relax}%
\providecommand \BibitemShut  [1]{\csname bibitem#1\endcsname}%
\let\auto@bib@innerbib\@empty
\bibitem [{\citenamefont {Heemskerk}\ \emph {et~al.}(2009)\citenamefont
  {Heemskerk}, \citenamefont {Penedones}, \citenamefont {Polchinski},\ and\
  \citenamefont {Sully}}]{Heemskerk:2009pn}%
  \BibitemOpen
  \bibfield  {author} {\bibinfo {author} {\bibfnamefont {I.}~\bibnamefont
  {Heemskerk}}, \bibinfo {author} {\bibfnamefont {J.}~\bibnamefont
  {Penedones}}, \bibinfo {author} {\bibfnamefont {J.}~\bibnamefont
  {Polchinski}}, \ and\ \bibinfo {author} {\bibfnamefont {J.}~\bibnamefont
  {Sully}},\ }\href {\doibase 10.1088/1126-6708/2009/10/079} {\bibfield
  {journal} {\bibinfo  {journal} {JHEP}\ }\textbf {\bibinfo {volume} {10}},\
  \bibinfo {pages} {079} (\bibinfo {year} {2009})},\ \Eprint
  {http://arxiv.org/abs/0907.0151} {arXiv:0907.0151 [hep-th]} \BibitemShut
  {NoStop}%
\bibitem [{\citenamefont {Fitzpatrick}\ and\ \citenamefont
  {Kaplan}(2012)}]{Fitzpatrick:2011dm}%
  \BibitemOpen
  \bibfield  {author} {\bibinfo {author} {\bibfnamefont {A.~L.}\ \bibnamefont
  {Fitzpatrick}}\ and\ \bibinfo {author} {\bibfnamefont {J.}~\bibnamefont
  {Kaplan}},\ }\href {\doibase 10.1007/JHEP10(2012)032} {\bibfield  {journal}
  {\bibinfo  {journal} {JHEP}\ }\textbf {\bibinfo {volume} {10}},\ \bibinfo
  {pages} {032} (\bibinfo {year} {2012})},\ \Eprint
  {http://arxiv.org/abs/1112.4845} {arXiv:1112.4845 [hep-th]} \BibitemShut
  {NoStop}%
\bibitem [{\citenamefont {Penedones}(2011)}]{Penedones:2010ue}%
  \BibitemOpen
  \bibfield  {author} {\bibinfo {author} {\bibfnamefont {J.}~\bibnamefont
  {Penedones}},\ }\href {\doibase 10.1007/JHEP03(2011)025} {\bibfield
  {journal} {\bibinfo  {journal} {JHEP}\ }\textbf {\bibinfo {volume} {03}},\
  \bibinfo {pages} {025} (\bibinfo {year} {2011})},\ \Eprint
  {http://arxiv.org/abs/1011.1485} {arXiv:1011.1485 [hep-th]} \BibitemShut
  {NoStop}%
\bibitem [{\citenamefont {Mueck}\ and\ \citenamefont
  {Viswanathan}(1998)}]{Muck:1998rr}%
  \BibitemOpen
  \bibfield  {author} {\bibinfo {author} {\bibfnamefont {W.}~\bibnamefont
  {Mueck}}\ and\ \bibinfo {author} {\bibfnamefont {K.~S.}\ \bibnamefont
  {Viswanathan}},\ }\href {\doibase 10.1103/PhysRevD.58.041901} {\bibfield
  {journal} {\bibinfo  {journal} {Phys. Rev.}\ }\textbf {\bibinfo {volume}
  {D58}},\ \bibinfo {pages} {041901} (\bibinfo {year} {1998})},\ \Eprint
  {http://arxiv.org/abs/hep-th/9804035} {arXiv:hep-th/9804035 [hep-th]}
  \BibitemShut {NoStop}%
\bibitem [{\citenamefont {Dolan}\ and\ \citenamefont
  {Osborn}(2001)}]{Dolan:2000ut}%
  \BibitemOpen
  \bibfield  {author} {\bibinfo {author} {\bibfnamefont {F.~A.}\ \bibnamefont
  {Dolan}}\ and\ \bibinfo {author} {\bibfnamefont {H.}~\bibnamefont {Osborn}},\
  }\href {\doibase 10.1016/S0550-3213(01)00013-X} {\bibfield  {journal}
  {\bibinfo  {journal} {Nucl. Phys.}\ }\textbf {\bibinfo {volume} {B599}},\
  \bibinfo {pages} {459} (\bibinfo {year} {2001})},\ \Eprint
  {http://arxiv.org/abs/hep-th/0011040} {arXiv:hep-th/0011040} \BibitemShut
  {NoStop}%
\bibitem [{\citenamefont {Bertan}\ \emph {et~al.}(2019)\citenamefont {Bertan},
  \citenamefont {Sachs},\ and\ \citenamefont {Skvortsov}}]{Bertan:2018afl}%
  \BibitemOpen
  \bibfield  {author} {\bibinfo {author} {\bibfnamefont {I.}~\bibnamefont
  {Bertan}}, \bibinfo {author} {\bibfnamefont {I.}~\bibnamefont {Sachs}}, \
  and\ \bibinfo {author} {\bibfnamefont {E.~D.}\ \bibnamefont {Skvortsov}},\
  }\href {\doibase 10.1007/JHEP02(2019)099} {\bibfield  {journal} {\bibinfo
  {journal} {JHEP}\ }\textbf {\bibinfo {volume} {02}},\ \bibinfo {pages} {099}
  (\bibinfo {year} {2019})},\ \Eprint {http://arxiv.org/abs/1810.00907}
  {arXiv:1810.00907 [hep-th]} \BibitemShut {NoStop}%
\bibitem [{\citenamefont {Bertan}\ and\ \citenamefont
  {Sachs}(2018)}]{Bertan:2018khc}%
  \BibitemOpen
  \bibfield  {author} {\bibinfo {author} {\bibfnamefont {I.}~\bibnamefont
  {Bertan}}\ and\ \bibinfo {author} {\bibfnamefont {I.}~\bibnamefont {Sachs}},\
  }\href {\doibase 10.1103/PhysRevLett.121.101601} {\bibfield  {journal}
  {\bibinfo  {journal} {Phys. Rev. Lett.}\ }\textbf {\bibinfo {volume} {121}},\
  \bibinfo {pages} {101601} (\bibinfo {year} {2018})},\ \Eprint
  {http://arxiv.org/abs/1804.01880} {arXiv:1804.01880 [hep-th]} \BibitemShut
  {NoStop}%
\bibitem [{\citenamefont {Heckelbacher}\ \emph {et~al.}(2022)\citenamefont
  {Heckelbacher}, \citenamefont {Sachs}, \citenamefont {Skvortsov},\ and\
  \citenamefont {Vanhove}}]{Heckelbacher:2022fbx}%
  \BibitemOpen
  \bibfield  {author} {\bibinfo {author} {\bibfnamefont {T.}~\bibnamefont
  {Heckelbacher}}, \bibinfo {author} {\bibfnamefont {I.}~\bibnamefont {Sachs}},
  \bibinfo {author} {\bibfnamefont {E.}~\bibnamefont {Skvortsov}}, \ and\
  \bibinfo {author} {\bibfnamefont {P.}~\bibnamefont {Vanhove}},\ }\href
  {\doibase 10.1007/JHEP08(2022)052} {\bibfield  {journal} {\bibinfo  {journal}
  {JHEP}\ }\textbf {\bibinfo {volume} {08}},\ \bibinfo {pages} {052} (\bibinfo
  {year} {2022})},\ \Eprint {http://arxiv.org/abs/2201.09626} {arXiv:2201.09626
  [hep-th]} \BibitemShut {NoStop}%
\bibitem [{\citenamefont {Moshe}\ and\ \citenamefont
  {Zinn-Justin}(2003)}]{Moshe:2003xn}%
  \BibitemOpen
  \bibfield  {author} {\bibinfo {author} {\bibfnamefont {M.}~\bibnamefont
  {Moshe}}\ and\ \bibinfo {author} {\bibfnamefont {J.}~\bibnamefont
  {Zinn-Justin}},\ }\href {\doibase 10.1016/S0370-1573(03)00263-1} {\bibfield
  {journal} {\bibinfo  {journal} {Phys. Rept.}\ }\textbf {\bibinfo {volume}
  {385}},\ \bibinfo {pages} {69} (\bibinfo {year} {2003})},\ \Eprint
  {http://arxiv.org/abs/hep-th/0306133} {arXiv:hep-th/0306133 [hep-th]}
  \BibitemShut {NoStop}%
\bibitem [{\citenamefont {Coleman}\ \emph {et~al.}(1974)\citenamefont
  {Coleman}, \citenamefont {Jackiw},\ and\ \citenamefont
  {Politzer}}]{Coleman:1974jh}%
  \BibitemOpen
  \bibfield  {author} {\bibinfo {author} {\bibfnamefont {S.~R.}\ \bibnamefont
  {Coleman}}, \bibinfo {author} {\bibfnamefont {R.}~\bibnamefont {Jackiw}}, \
  and\ \bibinfo {author} {\bibfnamefont {H.~D.}\ \bibnamefont {Politzer}},\
  }\href {\doibase 10.1103/PhysRevD.10.2491} {\bibfield  {journal} {\bibinfo
  {journal} {Phys. Rev. D}\ }\textbf {\bibinfo {volume} {10}},\ \bibinfo
  {pages} {2491} (\bibinfo {year} {1974})}\BibitemShut {NoStop}%
\bibitem [{\citenamefont {Meltzer}\ \emph {et~al.}(2020)\citenamefont
  {Meltzer}, \citenamefont {Perlmutter},\ and\ \citenamefont
  {Sivaramakrishnan}}]{Meltzer:2019nbs}%
  \BibitemOpen
  \bibfield  {author} {\bibinfo {author} {\bibfnamefont {D.}~\bibnamefont
  {Meltzer}}, \bibinfo {author} {\bibfnamefont {E.}~\bibnamefont {Perlmutter}},
  \ and\ \bibinfo {author} {\bibfnamefont {A.}~\bibnamefont
  {Sivaramakrishnan}},\ }\href {\doibase 10.1007/JHEP03(2020)061} {\bibfield
  {journal} {\bibinfo  {journal} {JHEP}\ }\textbf {\bibinfo {volume} {03}},\
  \bibinfo {pages} {061} (\bibinfo {year} {2020})},\ \Eprint
  {http://arxiv.org/abs/1912.09521} {arXiv:1912.09521 [hep-th]} \BibitemShut
  {NoStop}%
\bibitem [{\citenamefont {Carmi}(2020)}]{Carmi:2019ocp}%
  \BibitemOpen
  \bibfield  {author} {\bibinfo {author} {\bibfnamefont {D.}~\bibnamefont
  {Carmi}},\ }\href {\doibase 10.1007/JHEP06(2020)049} {\bibfield  {journal}
  {\bibinfo  {journal} {JHEP}\ }\textbf {\bibinfo {volume} {06}},\ \bibinfo
  {pages} {049} (\bibinfo {year} {2020})},\ \Eprint
  {http://arxiv.org/abs/1910.14340} {arXiv:1910.14340 [hep-th]} \BibitemShut
  {NoStop}%
\bibitem [{\citenamefont {Costa}\ \emph {et~al.}(2014)\citenamefont {Costa},
  \citenamefont {Goncalves},\ and\ \citenamefont {Penedones}}]{Costa:2014kfa}%
  \BibitemOpen
  \bibfield  {author} {\bibinfo {author} {\bibfnamefont {M.~S.}\ \bibnamefont
  {Costa}}, \bibinfo {author} {\bibfnamefont {V.}~\bibnamefont {Goncalves}}, \
  and\ \bibinfo {author} {\bibfnamefont {J.}~\bibnamefont {Penedones}},\ }\href
  {\doibase 10.1007/JHEP09(2014)064} {\bibfield  {journal} {\bibinfo  {journal}
  {JHEP}\ }\textbf {\bibinfo {volume} {09}},\ \bibinfo {pages} {064} (\bibinfo
  {year} {2014})},\ \Eprint {http://arxiv.org/abs/1404.5625} {arXiv:1404.5625
  [hep-th]} \BibitemShut {NoStop}%
\bibitem [{\citenamefont {Giombi}\ \emph {et~al.}(2013)\citenamefont {Giombi},
  \citenamefont {Klebanov}, \citenamefont {Pufu}, \citenamefont {Safdi},\ and\
  \citenamefont {Tarnopolsky}}]{Giombi:2013yva}%
  \BibitemOpen
  \bibfield  {author} {\bibinfo {author} {\bibfnamefont {S.}~\bibnamefont
  {Giombi}}, \bibinfo {author} {\bibfnamefont {I.~R.}\ \bibnamefont
  {Klebanov}}, \bibinfo {author} {\bibfnamefont {S.~S.}\ \bibnamefont {Pufu}},
  \bibinfo {author} {\bibfnamefont {B.~R.}\ \bibnamefont {Safdi}}, \ and\
  \bibinfo {author} {\bibfnamefont {G.}~\bibnamefont {Tarnopolsky}},\ }\href
  {\doibase 10.1007/JHEP10(2013)016} {\bibfield  {journal} {\bibinfo  {journal}
  {JHEP}\ }\textbf {\bibinfo {volume} {10}},\ \bibinfo {pages} {016} (\bibinfo
  {year} {2013})},\ \Eprint {http://arxiv.org/abs/1306.5242} {arXiv:1306.5242
  [hep-th]} \BibitemShut {NoStop}%
\bibitem [{\citenamefont {Ponomarev}(2020)}]{Ponomarev:2019ofr}%
  \BibitemOpen
  \bibfield  {author} {\bibinfo {author} {\bibfnamefont {D.}~\bibnamefont
  {Ponomarev}},\ }\href {\doibase 10.1007/JHEP01(2020)154} {\bibfield
  {journal} {\bibinfo  {journal} {JHEP}\ }\textbf {\bibinfo {volume} {01}},\
  \bibinfo {pages} {154} (\bibinfo {year} {2020})},\ \Eprint
  {http://arxiv.org/abs/1908.03974} {arXiv:1908.03974 [hep-th]} \BibitemShut
  {NoStop}%
\bibitem [{\citenamefont {Carmi}\ \emph {et~al.}(2019)\citenamefont {Carmi},
  \citenamefont {Di~Pietro},\ and\ \citenamefont {Komatsu}}]{Carmi:2018qzm}%
  \BibitemOpen
  \bibfield  {author} {\bibinfo {author} {\bibfnamefont {D.}~\bibnamefont
  {Carmi}}, \bibinfo {author} {\bibfnamefont {L.}~\bibnamefont {Di~Pietro}}, \
  and\ \bibinfo {author} {\bibfnamefont {S.}~\bibnamefont {Komatsu}},\ }\href
  {\doibase 10.1007/JHEP01(2019)200} {\bibfield  {journal} {\bibinfo  {journal}
  {JHEP}\ }\textbf {\bibinfo {volume} {01}},\ \bibinfo {pages} {200} (\bibinfo
  {year} {2019})},\ \Eprint {http://arxiv.org/abs/1810.04185} {arXiv:1810.04185
  [hep-th]} \BibitemShut {NoStop}%
\bibitem [{\citenamefont {Landau}\ \emph {et~al.}(1954)\citenamefont {Landau},
  \citenamefont {Abrikosov},\ and\ \citenamefont
  {Khalatnikov}}]{Landau:1954cxv}%
  \BibitemOpen
  \bibfield  {author} {\bibinfo {author} {\bibfnamefont {L.~D.}\ \bibnamefont
  {Landau}}, \bibinfo {author} {\bibfnamefont {A.~A.}\ \bibnamefont
  {Abrikosov}}, \ and\ \bibinfo {author} {\bibfnamefont {I.~M.}\ \bibnamefont
  {Khalatnikov}},\ }\href {\doibase 10.1016/b978-0-08-010586-4.50085-7}
  {\bibfield  {journal} {\bibinfo  {journal} {Dokl. Akad. Nauk SSSR}\ }\textbf
  {\bibinfo {volume} {95}},\ \bibinfo {pages} {1177} (\bibinfo {year}
  {1954})}\BibitemShut {NoStop}%
\bibitem [{\citenamefont {Caron-Huot}(2017)}]{Caron-Huot:2017vep}%
  \BibitemOpen
  \bibfield  {author} {\bibinfo {author} {\bibfnamefont {S.}~\bibnamefont
  {Caron-Huot}},\ }\href {\doibase 10.1007/JHEP09(2017)078} {\bibfield
  {journal} {\bibinfo  {journal} {JHEP}\ }\textbf {\bibinfo {volume} {09}},\
  \bibinfo {pages} {078} (\bibinfo {year} {2017})},\ \Eprint
  {http://arxiv.org/abs/1703.00278} {arXiv:1703.00278 [hep-th]} \BibitemShut
  {NoStop}%
\bibitem [{\citenamefont {Heckelbacher}\ and\ \citenamefont
  {Sachs}(2021)}]{Heckelbacher:2020nue}%
  \BibitemOpen
  \bibfield  {author} {\bibinfo {author} {\bibfnamefont {T.}~\bibnamefont
  {Heckelbacher}}\ and\ \bibinfo {author} {\bibfnamefont {I.}~\bibnamefont
  {Sachs}},\ }\href {\doibase 10.1007/JHEP02(2021)151} {\bibfield  {journal}
  {\bibinfo  {journal} {JHEP}\ }\textbf {\bibinfo {volume} {02}},\ \bibinfo
  {pages} {151} (\bibinfo {year} {2021})},\ \Eprint
  {http://arxiv.org/abs/2009.06511} {arXiv:2009.06511 [hep-th]} \BibitemShut
  {NoStop}%
\end{thebibliography}
\end{document}